\documentclass[aps,prb,floatfix,twocolumn,showpacs]{revtex4}

\usepackage{amsmath}
\usepackage{amsfonts}
\usepackage{amssymb}
\usepackage{graphicx}
\usepackage{bm}
\usepackage{color}
\usepackage[hypertex]{hyperref}

\newcommand{\bra}[1] {| #1 \rangle}
\newcommand{\ket}[1] {\langle #1| }

\def\stacksymbols #1#2#3#4{\def\theguybelow{#2}
    \def\verticalposition{\lower#3pt}
    \def\spacingwithinsymbol{\baselineskip0pt\lineskip#4pt}
    \mathrel{\mathpalette\intermediary#1}}
\def\intermediary#1#2{\verticalposition\vbox{\spacingwithinsymbol
      \everycr={}\tabskip0pt
      \halign{$\mathsurround0pt#1\hfil##\hfil$\crcr#2\crcr
               \theguybelow\crcr}}}

\def\gapproxeq{\stacksymbols{>}{\sim}{3}{.5}}

\begin{document}
\title{
Topological phases and delocalization of 
quantum walks in random environments
}
\author{Hideaki Obuse}
\altaffiliation{Present address: Institut f\"ur Nanotechnologie,
Karlsruhe Institute of Technology, 76021 Karlsruhe, Germany}
\affiliation{Department of Physics,
Kyoto University, Kyoto 606-8502, Japan}
\author{Norio Kawakami}
\affiliation{Department of Physics,
Kyoto University, Kyoto 606-8502, Japan}

\begin{abstract}
We investigate one-dimensional (1D) discrete time quantum walks (QWs)
 with spatially or temporally random defects as a consequence of
 interactions with random environments. We focus on the QWs with chiral
 symmetry in a topological phase, and reveal that chiral symmetry
 together with bipartite nature of the QWs brings about intriguing
 behaviors such as coexistence of topologically protected edge states at
 zero energy and Anderson transitions in the 1D chiral class at {\it
 non-zero} energy in their dynamics.  Contrary to the previous studies,
 therefore, the spatially disordered QWs can avoid complete localization
 due to the Anderson transition.  It is further confirmed that the edge
 states are robust for spatial disorder but not for temporal disorder.
\end{abstract}


\date{\today}

\maketitle

\section{introduction}
A quantum walk (QW)\cite{Aharonov93} describes quantum mechanical time
evolution of particles, which is identified as a random walk when the
system is brought to the classical limit.  The QW may provide a unique
avenue to realize quantum computation since the QW can be applied for
efficient algorithms of quantum computation.\cite{Kempe03,Ambainis03}
Among several kinds of QWs, the two-state discrete-time QW in one
dimension (1D) has been intensively investigated due to its simple
formalism. More remarkably, it has been experimentally realized in
various systems, such as cold atoms\cite{Karski09}, trapped
ions\cite{Schmitz09,Zahringer10}, and
photons.\cite{Schreiber10,Broome10,Schreiber11}
The discrete-time QW is described by two basic operators.  The coin
operator $C$ defined as
\begin{eqnarray}
C&\equiv&
\left(
\begin{array}{cc}
\cos \theta& -\sin \theta \\
 \sin \theta &  \cos \theta
\end{array}
\right),
\label{eq:coin}
\end{eqnarray}
acts on two internal states, right and left
walkers $\bra{R}$ and $\bra{L}$, respectively.
The shift operator $S$ is defined as
\begin{equation}
S\equiv
\sum_{n=-N/2}^{N/2-1}
\bigl(
\bra{n+1}\ket{n}\otimes\bra{R}\ket{R}+
\bra{n-1}\ket{n}\otimes\bra{L}\ket{L}
\bigr),
\label{eq:shift}
\end{equation}
so that each walker moves from a position $n$ to its corresponding
neighbor $n\pm 1$. Here, $N$ represents the system length.  The time
evolution operator $U$ is built up from these two operators and the
corresponding Hamiltonian $H$ is defined through
\begin{equation}
\textstyle
U\equiv S \Bigl( \sum_n \bra{n}\ket{n}\otimes C \Bigr)
=e^{i H \delta t},
\label{eq:time-evolution}
\end{equation}
where $\delta t$ is a unit of time and we hereafter set $\delta t=1$.
Thus, a state at time $t$, $\bra{\psi(t)}$, is given by
$\bra{\psi(t)}=U^t\bra{\psi(0)}$ starting from an initial state
$\bra{\psi(0)}$.

It is known that the Hamiltonian $H$ described by Eqs.\ (\ref{eq:coin})
- (\ref{eq:time-evolution}) is equivalent to the Dirac equation in the
continuum limit.\cite{Meyer96,Strauch06} Recently, symmetries of the
Hamiltonian, which are relevant to classify topological phases
\cite{Schnyder08,Kitaev09}, are examined in Ref.\
\onlinecite{Kitagawa10}, and it is found that the QW described by Eqs.\
(\ref{eq:coin})-(\ref{eq:time-evolution}), which has been realized in
many experiments
\cite{Karski09,Schmitz09,Zahringer10,Schreiber10,Broome10}, possesses
chiral symmetry.  Ref.\ \onlinecite{Kitagawa10} clarified that the Dirac
equation derived from the 1D QW with chiral symmetry gives a finite
Berry phase and thus generates edge states near boundaries, since 1D
chiral classes can be characterized by an integer topological
number\cite{Schnyder08,Kitaev09}.  The edge states in the topological
phase would be also useful for topological quantum
computation\cite{Kitaev01,Alicea11,Lehman11}, if the system is described
by Majorana fermions, which is currently a subject of intensive studies
in condensed matter physics.

Understanding effects of the interaction with environments giving rise
to decoherence is a key issue to realize the quantum computation.
\cite{Kendon07} To this end, there are
theoretical\cite{Mackay02,Brun03,Kosik06,Abal08,Konno09,
Joye10,Ahlbrecht11} and
experimental\cite{Karski09,Schreiber10,Broome10,Schreiber11} studies of
discrete-time QWs taking account of spatial and/or temporal disorders.
It is found that Anderson localizations occur for QWs with spatial
disorder while QWs with temporal disorder approach to classical random
walks.

In this work, we investigate two-state discrete time 1D QWs described by
Eqs.(\ref{eq:coin})-(\ref{eq:time-evolution}), which belong to the
chiral orthogonal class, interacting with spatially or temporally random
defects.  Contrary to the previous studies, we reveal that the QWs with
spatial disorder exhibit delocalized behaviors.  This remarkable
conclusion is drawn by extensive numerical calculations, and is further
ensured by symmetry arguments for the 1D chiral class.  We also find
that the edge states in the topological phase are robust for spatial
disorder but not for temporal disorder.

This paper is organized as follows.  In the Sec.\ II, we give general
remarks on QWs without any disorder and demonstrate that the edge states
are induced by introducing a reflecting coin operator. Our main results
for the spatially or temporally disordered QWs are presented in Sec.\
III. Section IV is devoted to discussions and summary.

\section{Quantum Walks with reflecting coin operators}

\begin{figure}[t]
\includegraphics[width=8cm]{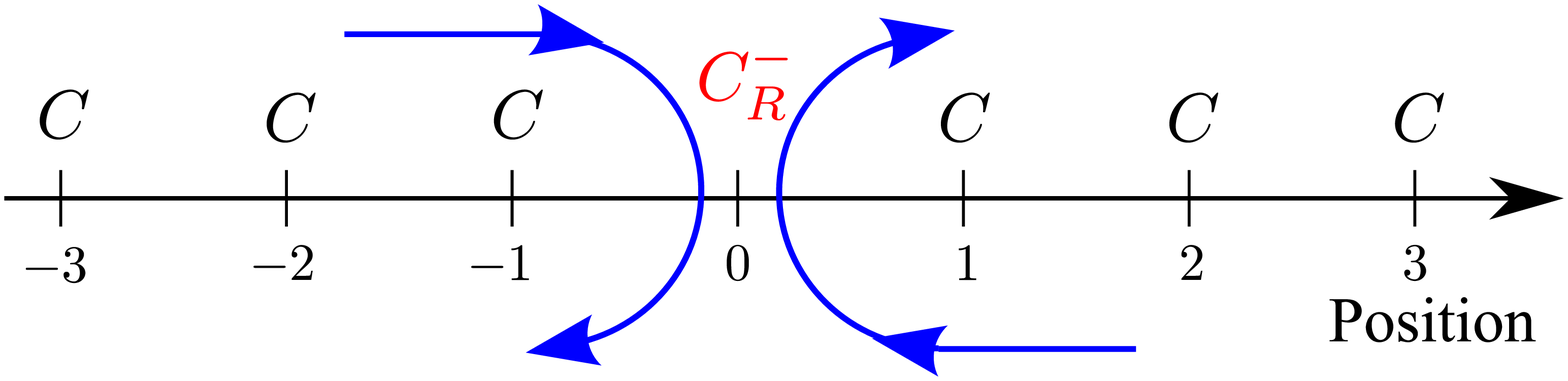}
\caption{
(Color online)
The QW with a reflecting coin operator at the origin.
}
\label{fig:RBC}
\end{figure}

Let us first explain eigenstates of the QW.  Eigenenergy
$\omega_\lambda$ of the QW, which has periodicity of $2\pi$, is defined
through the eigenvalue of the time-evolution operator by $U
\bra{\psi_\lambda} = e^{i \omega_\lambda} \bra{\psi_\lambda}$, where
$\bra{\psi_\lambda}$ is the corresponding eigenstate.  Hereafter, the
subscript $\lambda$ is written only when we need to emphasize.  The
dispersion relation resulting from Eqs.\
(\ref{eq:coin})-(\ref{eq:time-evolution}) is given by
$ \cos(\omega) = \cos(k)\cos(\theta),
$
where $k$ is a wave number.\cite{Strauch06,Kitagawa10} Since the energy
gap should exist to support edge states in topological phases, edge
states of the QW are able to appear when $\theta \ne 0,\pi$.

To generate the edge states, a boundary should be prepared for the QW.
While a split-step method is proposed in Ref.\ \onlinecite{Kitagawa10},
we employ a simpler method to make the boundary in terms of a reflecting
coin operator defined as
\begin{eqnarray}
C_{R}^{\pm}&\equiv&
\left(
\begin{array}{cc}
0& \mp 1 \\
\pm 1 &  0
\end{array}
\right), 
\end{eqnarray}
so that a right walker is changed to a left walker and vice
versa. Thereby, $C_R^{\pm}$ realizes a hard wall boundary. The sign of
the reflecting coin operator determines the presence or absence of edge
states.\cite{Oka05,Chisaki10} In case of the Hadamard walks
($\theta=\pi/4$), the edge states appear if $C_R^{-}$ is introduced at
the boundary.

\begin{figure}[t]
\includegraphics[width=8.5cm]{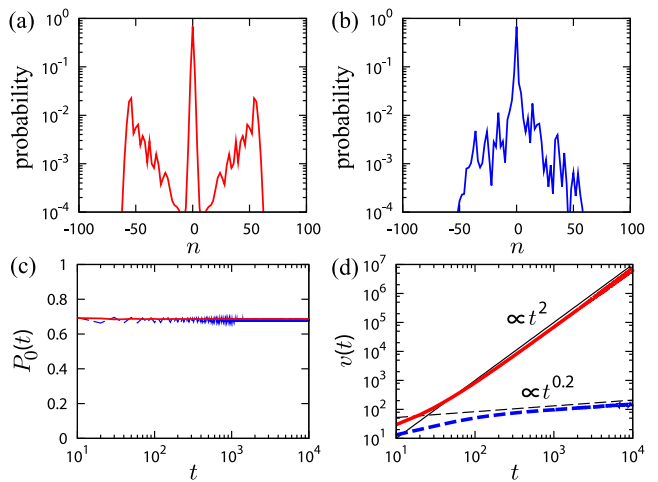} 
\caption{ (Color online)
Probability distributions at $t=80$ of the QWs with $\delta \theta_s =0$
(a) and $\pi/4$ (b) with $C_R^-$ for a single sample.  $t$ dependence of
the survival probability $P_0(t)$ (c) and the variance $v(t)$ (d).  In
(c) and (d), the solid and dashed curves represent the QWs with $\delta
\theta_s=0$ and $\pi/4$, respectively.  The two thin lines in (d)
indicate power law behaviors.  The number of samples is $10^4$ in case
of disordered QWs.  } \label{fig:probability}
\end{figure}

We now consider the Hadamard walks with $C_R^{-}$ at the origin as shown
in Fig.\ \ref{fig:RBC}.  The initial state is prepared as
\begin{equation}
\bra{\psi(0)}\equiv\bra{0}\otimes(\bra{R}+i\bra{L})/\sqrt{2}
\label{eq:initial}
\end{equation}
throughout the paper, so that the right and left walkers do not mix with
each other due to real-number elements of the coin operator and then the
probability distribution becomes symmetric in the absence of spatial
disorder.  Figure \ref{fig:probability} (a) shows the probability
distribution $P_n(t)$ at $t=80$, which is defined as
\begin{equation}
P_n(t) \equiv \sum_{\sigma=R,L}|\ket{n}\otimes\ket{\sigma}\psi(t)\rangle|^2.
\end{equation}
We find a sharp peak at the origin due to edge states in the topological
phase and two smaller peaks near $n\sim \pm 50$ commonly observed in
QWs.\cite{Kempe03,Ambainis03}.  Note that these observations are
consistent with the recently derived limit measure\cite{Chisaki10} at $t
\rightarrow \infty$ for the 1D QW on a half line. The time dependence of
the survival probability
\begin{equation}
P_0(t)\equiv P_n(t)|_{n=0}
\end{equation}
and the position variance of QWs
\begin{equation}
v(t) \equiv 
\langle\psi(t)|n^2|\psi(t)\rangle
- 
[
\langle\psi(t)|n|\psi(t)\rangle
]^2
\end{equation}
are shown by solid curves in Fig.\ \ref{fig:probability} (c) and (d),
respectively.  We confirm that the edge states are robust for time
evolution since $P_0(t)$ keeps its value unchanged and $v(t)$ is
proportional to $t^2$, which agrees with ordinary
QWs.\cite{Kempe03,Ambainis03}

\section{Disordered QWs}

Next, we consider disordered QWs.  We assume that disorder is introduced
via fluctuations of $\theta$ in Eq. (\ref{eq:coin}), which do not break
chiral symmetry.  To this end, we redefine $\theta$ so that $\theta$ is
randomly distributed over positions $n$ or time $t$ as
\begin{equation*}
 \theta \in
[\overline \theta - \delta \theta_{s(t)}/2 : \overline \theta 
+\delta \theta_{s(t)}/2],
\end{equation*}
where $\overline \theta$ represents the mean value of distributed
$\theta$ and $\delta \theta_{s(t)}$ indicates the strength of spatial
(temporal) disorder.  We hereafter restrict our attention to QWs with
$\overline \theta=\pi/4$ and with either spatial or temporal disorder.

\subsection{spatially disordered QWs}

Let us begin with the spatial disorder.  The probability distribution
$P_n(t)$ for spatially disordered QWs with $C_R^{-}$ after $80$ time
steps is shown in Fig.\ \ref{fig:probability} (b).  The peak at the
origin due to the edge states seems to be robust for the spatial
disorder.  The survival probability $P_0(t)$ indeed ensures this point:
the profile of $P_0(t)$ is almost unchanged from that for the clean QW.
On the other hand, the outer two smaller peaks existing in Fig.\
\ref{fig:probability} (a) disappear in Fig.\ \ref{fig:probability} (b),
implying that the QWs suffer from the spatial disorder except their edge
states.  In this case, the Anderson localization would be important as
studied in the previous works.
\cite{Schreiber11,Joye10,Ahlbrecht11} If the Anderson
localization completely dominates dynamics of QWs, the variance $v(t)$
should become a constant after many time steps.  However, $v(t)$ for the
spatially disordered QWs [Fig.\ \ref{fig:probability} (d)] exhibits a
power law behavior with a smaller exponent, suggesting that the QWs show
anomalous diffusion even in the presence of spatial disorder.

\begin{figure}[t]
\includegraphics[width=8.5cm]{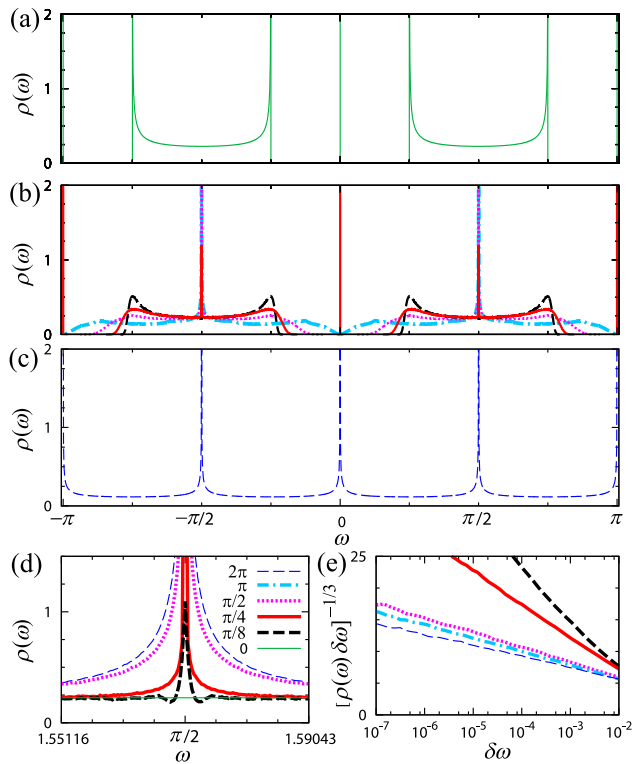} 
\caption{ (Color online) DOS
$\rho(\omega)$ for the QWs with $C_R^-$ for $\delta \theta_s=0$ (a),
$\delta \theta_s =\pi/8,\pi/4,\pi/2$, and $\pi$ (b), and $2\pi$ (c).
$N=500$ and the number of samples is $10^5$ for each $\delta \theta_s
(\ne0)$.  (d) The enlargement of DOS around $\omega=\pi/2$ for QWs with
various $\delta \theta_s$.  (e) The scaling collapses into Eq.\
(\ref{eq:dos}) for the DOS near $\omega=\pi/2$ for QWs of the larger
system $N=10^4$ with $10^4$ samples.  The labels in (d) represent values
of $\delta \theta_s$.  In calculations of DOS, the periodic boundary
 conditions are imposed on both edges of the 1D QW.
} \label{fig:dos}
\end{figure}

To address the above point, we look into the density of states (DOS),
$\rho(\omega)$, which provides us with a clear-cut interpretation of the
seemingly peculiar dynamics of the QWs.  Note that a state
$\bra{\psi(t)}$ is considered as the superposition of all the
eigenstates because we employ a delta-function like initial state
$\bra{\psi(0)}$.  The DOS for the clean QW with translational symmetry
is given as
\begin{equation}
\rho_c(\omega) = 
\frac{\sin \omega}{2 \pi \sqrt{\cos^2 \theta - \cos^2 \omega }}
\quad 
\text{for } \cos^2 \omega < \cos^2 \theta.
\label{eq:dos_clean}
\end{equation}
When $\cos^2 \omega = \cos^2 \theta$ at $\theta \ne 0,\pi$,
$\rho_c(\omega)$ diverges due to the van Hove singularity.  For the QW
with $C_R^{-}$, we find that the DOS consists of $\rho_c(\omega)$ and
doubly degenerate states at $\omega=0,\pi$ as shown in Fig.\
\ref{fig:dos} (a).  According to the nature of $\bra{\psi_\lambda}$ for
$\omega_\lambda=0,\pi$, these states are identified as the edge states
in the topological phase.  In the clean QW, therefore, the edge states
appear not only at $\omega=0$ but also at $\omega=\pi$.  With increasing
$\delta \theta_s$ [Fig.\ \ref{fig:dos}(b)], the divergences in the DOS
at $\omega=\pm \pi/4$ and $\pm 3\pi/4$ are rounded since the van Hove
singularity is not well defined for disordered systems.  Furthermore, we
notice that the energy gaps around $\omega=0,\pi$ are reduced while the
delta functions due to edge states robustly remain.

Remarkably, there appear additional divergences in the DOS at
$\omega=\pm \pi/2$ for spatially disordered QWs [Fig.\ \ref{fig:dos}
(b)].  We give attention to the divergences in $\rho(\omega)$ at
$\omega=\pm \pi/2$ by showing the enlarged picture of DOS near
$\omega=\pi/2$ in Fig.\ \ref{fig:dos} (d).  We find that the divergence
becomes stronger with increasing $\delta \theta_s$ and the DOS for the
smallest disorder ($\delta \theta_s=\pi/8$) shows the oscillating
behavior.  This oscillation reminds us of the DOS derived from chiral
random matrix theories.\cite{NSV} Indeed, a system belonging to the 1D
chiral class is known to show an Anderson transition at {\it zero
energy}, which is accompanied by the divergence in the DOS.\cite{DTE}
Although the energy at which the DOS diverges is different from what we
observed for the QW, we claim that {\it the Anderson transition in the
1D chiral class occurs in the spatially disordered QW at $\omega=\pm
\pi/2$}. This is supported by the following symmetry arguments.
Chiral symmetry gives a constraint that eigenenergies appear in pairs of
opposite sign and this makes the zero energy states in chiral classes
singular.  In the QWs, chiral symmetry leads to the eigenvalues of $U$
appearing in pairs like $e^{\pm i \omega_\lambda}$.  In addition, since
$S$ in Eq.\ (\ref{eq:shift}) gives only nearest neighbor hopping, $U$ in
Eq.\ (\ref{eq:time-evolution}) has the bipartite structure and the
eigenvalues of $U$ also appear in pairs as $\pm e^{i
\omega_\lambda}.$\cite{note1} Thereby, there are four eigenenergies
related to each other,
\begin{equation*}
(\omega_\lambda,\ -\omega_\lambda,\ 
\omega_\lambda+\pi,\ -\omega_\lambda+\pi).
\end{equation*}
Redefining $\omega_\lambda$ as $\omega_\lambda=\omega_\lambda'+\pi/2$, these
eigenenergies are rewritten as 
\begin{equation*}
(\omega_\lambda'+\pi/2,\  -\omega_\lambda'-\pi/2,\ 
\omega_\lambda'+3\pi/2,\ -\omega_\lambda'+\pi/2).
\end{equation*}
One finds that the first and fourth (second and third) eigenenergies
appear in pairs with respect to $\omega=\pi/2$
$(-\pi/2)$\cite{Shikano10}, making $\omega=\pm \pi/2$ special energies
as in zero energy of ordinary chiral classes.  We therefore come to the
conclusion that the Anderson transition of the 1D chiral class should
occur at $\omega=\pm \pi/2$ even if there are energy gaps and the edge
states exist at $\omega=0,\pi$.

According to universality, the same critical behaviors near zero energy
studied with a tight-binding Hamiltonian in Ref.\ \onlinecite{DTE}
should be observed in the discrete time QW near $\omega=\pm
\pi/2$. Thereby, we write down the modified formulae for divergences of
the DOS $\rho$ and the localization length $\xi$, defined as the inverse
of Lyapunov exponents, at $\omega=\pm \pi/2$ as
\begin{eqnarray}
\rho(\omega) &=&
\rho_0 (\delta \omega \tau |\ln^3(\delta \omega \tau)|)^{-1},
\label{eq:dos}\\
 \xi(\omega) &=& \xi_0 |\ln(\delta \omega \tau)|, 
\label{eq:localization}
\end{eqnarray}
respectively, where $\delta \omega \equiv |\omega \mp \pi/2|$ represents
the distance from the critical points $\omega=\pm \pi/2$. 
$\tau$ denotes the mean free time and $\rho_0$ and $\xi_0$ are
constants. As shown in Fig.\ \ref{fig:dos} (e), we verify that the DOS
near $\omega=\pi/2$ obeys Eq.\ (\ref{eq:dos}) fairly well. 
\begin{figure}[t]
\includegraphics[width=6cm]{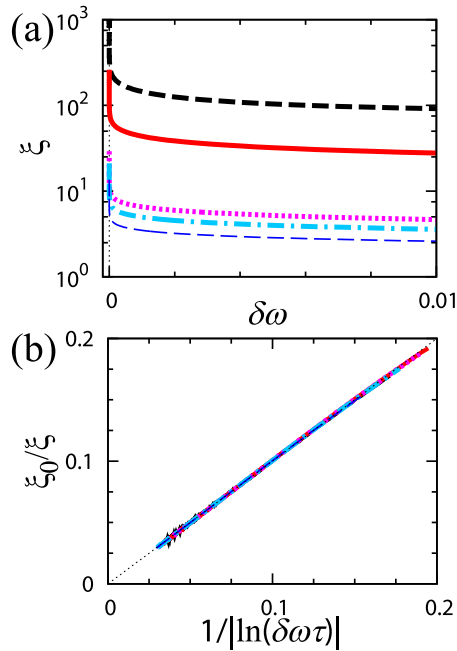} \caption{ (Color online) (a)
Dependence on $\delta \omega$ of the localization length $\xi$ of the
QWs with $\delta \theta_s=\pi/8,\pi/4,\pi/2,\pi,$ and $2\pi$ from the
top to the bottom.  (b) The scaling collapse into Eq.\
(\ref{eq:localization}) for $\xi$ in (a). The parameters $\xi_0$ and
$\tau$ are obtained by the fitting.  The dotted thin line represents
Eq.\ (\ref{eq:localization}).  } \label{fig:localization}
\end{figure}

We also calculate the localization length $\xi$ of the spatially
disordered QW near $\omega=\pi/2$ by using the transfer matrix
method.\cite{MacKinnon83,Joye10} By rearranging the wave function
amplitudes around the position $n$ of eigenvalue equation $U \bra{\psi} 
= e^{i \omega} \bra{\psi}$ , we obtain the iterative relation,
\begin{eqnarray}
\left(
\begin{array}{c}
\psi_{n+1,R}
\\
\psi_{n,L}
\end{array}
\right)
&=&
T_n
\left(
\begin{array}{c}
\psi_{n,R}
\label{eq:transfermat1}
\\
\psi_{n-1,L}
\end{array}
\right),
\\
T_n&=&
\left(
\begin{array}{cc}
e^{-i \omega}/ \cos \theta_n & -\tan \theta_n \\
-\tan \theta_n  & e^{i\omega}/ \cos\theta_n
\end{array}
\right) ,
\label{eq:transfermat2}
\end{eqnarray}
where $\psi_{n,\sigma=R,L}$ represents the wave function amplitude of
the right or left walker at the position $n$ and $\theta_n$ denotes
$\theta$ at the position $n$.  The huge system length $N=10^8$ allows us
to evaluate accurate numerical values of $\xi$.

Figure \ref{fig:localization} (a) shows the $\delta \omega$ dependence
of $\xi$ for various $\delta \theta_s$.  We confirm that $\xi$ for any
$\delta \theta_s$ shows the diverging behavior in the vicinity of
$\delta \omega=0$ while $\xi$ for $\delta \omega \ne 0$ is finite,
resulting in the Anderson localization.  Although the numerical
restriction of $\delta \omega \gapproxeq 10^{-16}$ prevents us to reach
$\xi \rightarrow \infty$ with accuracy, we confirm that $\xi$ follows
Eq.\ (\ref{eq:localization}) as shown in Fig.\ \ref{fig:localization}
(b).  This scaling collapse gives an evidence that $\xi$ diverges at
$\delta \omega=0$ for any $\delta \theta_s$.  In this case, the
transport property is anomalous by involving algebraic decays instead of
the exponential ones.\cite{Mudry99} Hence, the walkers can move around
through these delocalized states at $\omega=\pm \pi/2$ and then $v(t)$
indicates anomalous diffusion as shown in Fig.\ \ref{fig:probability}
(d).  Interestingly, three kinds of states (edge, localized, and
delocalized states) are simultaneously appear in dynamics of spatially
disordered QWs with $C_R^-$. This coexistence occurs not only for the
specific initial state, Eq.\ (\ref{eq:initial}), but also for general
delta function like initial states, so that dynamics of QWs is described
by the whole eigen states.  We mention that since the coin operator of
the previous works for the 1D spatially disordered QWs
\cite{Schreiber11,Konno09,Joye10,Ahlbrecht11} does not retain chiral
symmetry, the Anderson transition is not found and only the Anderson
localized states are confirmed.

It is known for zero energy states of chiral classes\cite{Ryu02} and
various topological insulators, such as the integer quantum Hall
transition, that topologically protected edge states stably exist unless
the bulk band gap is closed.  When a symmetry-preserving perturbation
collapses the bulk gap, the delocalized states which can be coupled with
the edge states so that the topological number is changed should appear
even in 1D disordered systems.  In other words, the existence of the
delocalized states as the Anderson delocalization is essential to
topological phases in disordered systems.  It is naturally expected,
then, that the 1D QW with chiral symmetry also follows a similar fashion
under the spatial disorder of $\theta$.

\begin{figure}[t]
\includegraphics[width=7cm]{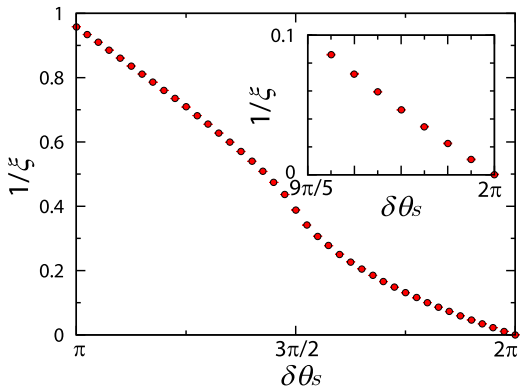} 
\caption{ (Color online)
$\delta \theta_s$ dependence of the inverse of the localization length at
$\omega=0$. The length of the system is $N=10^8$.  Inset: The
enlargement of $\xi^{-1}$ near $\delta \theta_s=2 \pi$.  }
\label{fig:localization_disorder}
\end{figure}

Here, we clarify how the strong spatial disorder $\delta \theta_s$
induces the delocalized states at $\omega=0,\pi$ as a consequence of the
gap closing.  Figure \ref{fig:dos} (b) indicates that the bulk gaps
around $\omega=0,\pi$ become smaller with increasing $\delta \theta_s$,
though there still remain the edge states at $\delta \theta_s = \pi$.
Unfortunately, it is difficult to find the gap closing from the DOS
since the finite system calculation keeps the mean level spacing finite.
Alternatively, we calculate the localization length $\xi$ at $\omega=0$
for the QW without $C_R^{-}$ for various $\delta \theta_s$ by using
Eqs.\ (\ref{eq:transfermat1}) and (\ref{eq:transfermat2}).  Note that
when the bulk gap around $\omega=0$ is finite, the localization length
is interpreted as of an exponentially decaying evanescent mode.  Figure
\ref{fig:localization_disorder} clearly shows that the localization
length $\xi$ at $\omega=0$ increases with increasing $\delta \theta_s$
and finally diverges at $\delta \theta_s=2 \pi$, where $\theta$ is fully
distributed in $2\pi$ radians.  This suggests that the bulk gap around
$\omega=0$ is closed.  Figure \ref{fig:dos} (c) shows the DOS
$\rho(\omega)$ at $\delta \theta_s=2 \pi$.  We confirm that, instead of
the delta functions of the edge states, the divergences in the DOS,
which are the same with those at $\omega=\pm \pi/2$, appear even at
$\omega=0,\pi$.  Therefore, the Anderson transitions occur at
$\omega=0,\pi$ as well as $\pm \pi/2$ at $\delta \theta_s = 2 \pi$ and
the coexistence of the edge states and the delocalized states does not
occur.

\subsection{temporally disordered QWs}

\begin{figure}[t]
\includegraphics[width=6cm]{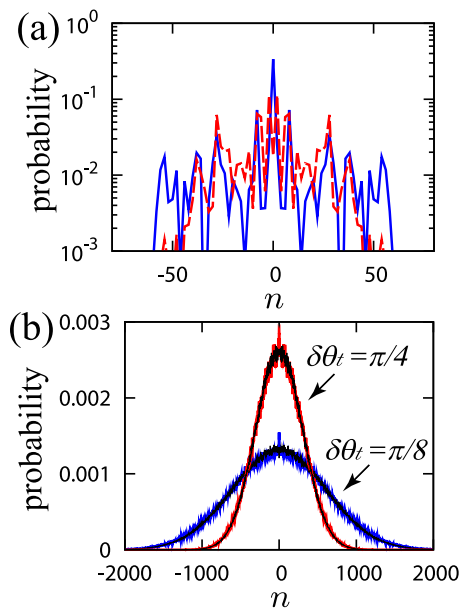} \caption{ (Color online)
Probability distributions at $t=80$ for a single sample (a) and sample
averaged probability distributions at $t=10^4$ (b) of the temporally
disordered QWs. The solid and dashed curves represent QWs with $\delta
\theta_t=\pi/8$ and $\pi/4$, respectively, and the thick and thin curves
distinguish the QWs with and without $C_R^-$, respectively. The number
of samples is $10^4$.  } \label{fig:dynamical}
\end{figure}

\begin{figure}[t]
\includegraphics[width=6cm]{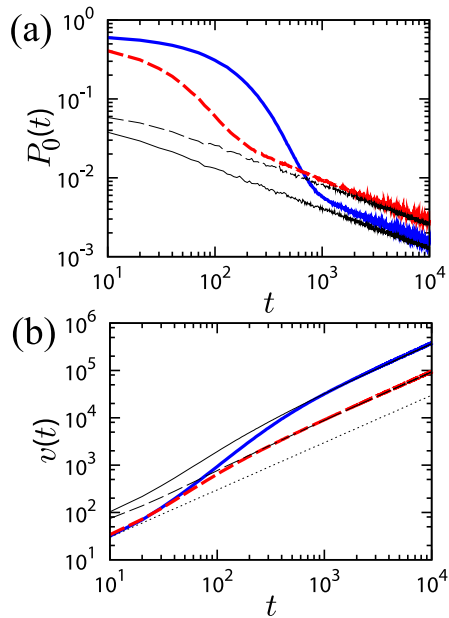} \caption{ (Color online) $t$
dependence of the survival probability $P_0(t)$ (a) and the variances
$v(t)$ (b).  The solid and dashed curves represent QWs with $\delta
\theta_t=\pi/8$ and $\pi/4$, respectively, and the thick and thin curves
distinguish the QWs with and without $C_R^-$, respectively.  The number
of samples is $10^4$.  The dotted line in (b) indicates $v(t)\propto t$.
} \label{fig:dynamical_2}
\end{figure}

Finally, we consider QWs with temporal disorder where $\theta$ in Eq.\
(\ref{eq:coin}) is shuffled in each time step.  It is known that, since
the temporal disorder gives rise to decoherence, the QWs approach to the
classical random walks where the probability distribution of QWs becomes
Gaussian and the position variance $v(t)$ is proportional to
$t$.\cite{Mackay02,Brun03,Kosik06,Abal08} We study how the edge states
of the QWs with $C_R^-$ are affected by the temporal disorder.  The
probability distribution $P_n(t)$ after $80$ time steps in Fig.\
\ref{fig:dynamical} (a) demonstrates that the peaks due to the edge
states are substantially reduced with increasing $\delta \theta_t$.  For
longer time steps $t=10^4$, the sample averaged $P_n(t)$ with temporal
disorder [Fig.\ \ref{fig:dynamical} (b)] become Gaussian, indicating
that the QWs approach toward the classical random walks.  The tiny peaks
at the origin, which would be the remnants of the edge states, are found
for the QWs with $C_R^{-}$.  In Fig.\ \ref{fig:dynamical_2} (a), the
survival probabilities $P_0(t)$ of QWs with $C_R^-$ are compared with
those without $C_R^-$.  We find that $P_0(t)$ of the former QWs behaves
differently from the latter during rather short time steps, while at
longer time steps they converge to the same curve, which means that the
edge states gradually disappear.  The position variance $v(t)$ in Fig.\
\ref{fig:dynamical_2} (b) increasing linearly in time for longer time
steps clarifies that the QWs with edge states are also transformed into
the classical random walks.

\section{Discussion and Summary}

In this work, we have focused only on discrete-time QWs.  Here some
comments are in order on the relevance of our results to continuous-time
QWs with spatial disorder.  While the 1D continuous-time QW with on-site
random potentials is studied previously\cite{Keating07,Yin08}, the
corresponding Hamiltonian does not retain chiral symmetry. We note that
the Hamiltonian of the 1D continuous-time QW possessing chiral symmetry
is the tight-binding model with only nearest neighbor random hopping
terms. Indeed, the latter model is consistent with the one studied in
Ref.\ \onlinecite{DTE}, and the Anderson transition is possible only at
zero energy in contrast to the discrete-time QWs studied in the present
work.  Therefore, the coexistence of edge states and the Anderson
transition is peculiar to discrete-time QWs.

We also mention that, although we have been concerned with the QWs with
$C_R^-$ here, our conclusion can directly be applied for the QWs without
$C_R^-$, except for the argument on edge states at $\omega=0,\pi$.

In summary, we have investigated two-state discrete time QWs belonging
to the 1D chiral orthogonal class in the presence of spatial or temporal
disorder.  We have elucidated that Anderson transitions in the 1D chiral
classes occur at $\omega=\pm \pi/2$ in the QWs with any strength of
spatial disorder and thereby the QWs can avoid the complete
localization.  We note that, while the currently available experiments
on discrete-time QWs can accomplish only a few ten time steps and it may
difficult to completely eliminate the temporal disorder giving rise to
decoherence in the spatially disordered QWs, the delocalization behavior
found in the present work will be observed when the technology of QWs
will be further developed in the application of quantum computer.

Furthermore, we have shown that the coexistence of edge, localized, and
delocalized states is realized in the time evolution of the spatially
disordered QW with $C_{R}^{-}$.  This characteristic nature of the QWs
is supported by chiral symmetry of $H$ and bipartite structures of $U$.
We predict that delocalization of QWs with spatial disorder should be
observed in a wide variety of 1D QWs \cite{Kitagawa10} belonging to the
other chiral classes and class D and DIII which are described by
Majorana fermions, since these universality classes also show the
Anderson transition in 1D.  \cite{Brouwer00,Titov01} The QWs for which
tuning system parameters and observing walkers' probabilities are
possible in experiments would provide an intriguing arena to study
topological phases for systems with defects and decoherence.

{\it Note:} After submitting the manuscript, we found a preprint
arXiv:1105.5334 \cite{Kitagawa11} in which the edge states of QWs are
observed in experiments.

\section{Acknowledgements}

We acknowledge fruitful discussions with T.\ Oka.  H.O. is supported by
Grant-in-Aid for JSPS for Young Scientists, and N.K. by KAKENHI
(Nos. 21740232, 20104010) and JSPS through the ``Funding Program for
World-Leading Innovative R\&D on Science and Technology (FIRST Program).

\end{document}